# Metáforas científicas no discurso jornalístico

# (Scientific Metaphors in the journalistic discourse)

Osame Kinouchi
Angélica A. Mandrá
Laboratório de Divulgação Científica e Cientometria
DFM – FFCLRP – USP

Email: osame@ffclrp.usp.br

**Resumo**: A educação e divulgação científicas não apenas ampliam o vocabulário e o repertório de conceitos científicos de uma população mas, ao mesmo tempo, promovem a difusão de certas metáforas conceituais e cognitivas. Neste trabalho, fazemos algumas hipóteses sobre este processo, propondo uma classificação em termos de metáforas visíveis, invisíveis, básicas e derivadas. Focalizamos nossa atenção em metáforas de física clássica e contemporânea aplicadas a fenômenos psicológicos e sócio-econômicos, e estudamos dois casos exemplares através do exame exaustivo do conteúdo online de grandes portais jornalísticos brasileiros. Finalmente, apresentamos implicações e sugestões da teoria de metáforas cognitivas para o processo de educação e divulgação científicas.

**Abstract**: Scientific education and divulgation not only amplify people's vocabulary and repertory of scientific concepts but, at the same time, promote the diffusion of certain conceptual and cognitive metaphors. Here we make some hypothesis about this process, proposing a classification in terms of visible, invisible, basic and derived metaphors. We focus our attention in contemporary and classical physics metaphors applied to psychological and socio-economical phenomena, and we study two exemplar cases through an exhaustive exam of the online content of large Brazilian journalistic portals. Finally, we present implications and suggestions from the cognitive metaphor theory for the scientific education and divulgation process.

*Good mathematicians see analogies between theorems or theories; the very best ones see analogies between analogies* – Stefan Banach, citado por S. M. Ulam.

## 1 Introdução

A linguagem coloquial, o padrão jornalístico, a norma culta e a linguagem científica não constituem universos estanques, mas interagem de maneiras as mais diversas e por vezes imprevisíveis. Ao longo de sua vida, estudantes e cidadãos leigos irão tomar contato, em maior ou menor grau, com esses diferentes universos lingüísticos. Neste trabalho faremos algumas considerações sobre como certos conceitos científicos acabam por se transferir para a linguagem comum através de um uso metafórico dos mesmos e que implicações isso pode ter para as atividades de divulgação científica e educação científica formal.

Embora a difusão de conceitos e termos científicos seja um processo complexo e multifacetado mediado pela educação formal, pela cultura popular (desenhos animados, cinema, videogames, quadrinhos, música pop etc.), pela mídia tradicional (jornalismo televisivo, documentários, jornais, revistas etc.) e pelas novas mídias (sites, blogs etc.), centraremos nossa atenção na questão da difusão de certos conceitos de física e astronomia, ou pelo menos do vocabulário a eles associado, através de metáforas presentes em textos jornalísticos voltados para a análise política e econômica.

Neste trabalho, o termo "metáfora" será usado em um sentido cognitivo em vez de apenas no sentido linguístico. Adotaremos a posição da teoria cognitiva de Lakoff e Johnson para os quais a essência da metáfora é "compreender e experimentar um tipo de coisa em termos de outra", um mapeamento entre um domínio fonte concreto e um domínio alvo abstrato [1,2]. Embora conscientes da existência de outras teorias da metáfora em lingüística, nossa preocupação é chamar a atenção para um fenômeno pouco estudado, mas potencialmente relevante para o ensino e divulgação científicos, em vez de atingir uma precisão lingüística prematura.

O foco deste trabalho em metáforas jornalísticas se justifica por motivos sociais e metodológicos. Em termos sociais, o jornalismo tem papel importante na manutenção de certos padrões lingüísticos e continua a ser um dos principais agentes de formação de opinião pública. Seja escrito por jornalistas científicos ou não, o texto jornalístico permanece como importante intermediário entre a comunidade acadêmico-científica e o grande público. Em termos metodológicos, a existência de grandes portais jornalísticos com ferramentas de busca por palavras-chave permite uma amostragem relativamente completa do uso de termos e metáforas científicas nesses ambientes.

O artigo é organizado da seguinte forma. Na seção 2, discutimos a definição de metáfora científica a ser usada neste trabalho e propomos uma classificação preliminar em termos de metáforas visíveis, invisíveis, básicas e derivadas. Na seção 3, apresentamos alguns exemplos de metáforas científicas originárias da geometria e física, examinadas segundo a classificação proposta, com ênfase em metáforas aplicadas a fatos sociais e econômicos. Um estudo de caso um pouco mais quantitativo de metáforas jornalísticas que utilizam termos de física e astronomia é apresentado na seção 4. Na seção 5, propomos que a educação e a divulgação científicas, para além da transmissão literal de conteúdos científicos, desempenham um papel relevante na ampliação do vocabulário de metáforas conceituais à disposição de uma população, aumentando seu poder de descrição, expressão e mesmo cognição de fatos políticos, econômicos e sociais. Em nossas conclusões, propomos que o exame das metáforas científicas na linguagem jornalística e coloquial poderia abrir novas perspectivas sobre a questão da relevância social da aprendizagem formal e informal de ciências.

## 2 Metáforas comuns e metáforas cientificamente inspiradas

Nas teorias cognitivas da metáfora [1-3] faz-se a distinção entre uma expressão metafórica (figura estilística de linguagem, por exemplo, "a máquina do Estado") e uma metáfora conceitual que configura, afeta e limita a própria cognição ("o Estado é um tipo de máquina"). Por simplicidade, usaremos neste trabalho o termo genérico "metáfora" em ambos os casos, qualificando o seu sentido apenas quando necessário.

### 2.1 Metáforas comuns e linguagem científica

Na literatura em ensino de física já existe uma série de trabalhos sobre o uso de metáforas e analogias como instrumentos pedagógicos [4-9]. Por exemplo, é possível fazer uma analogia (parcial) entre corrente elétrica em um fio condutor e corrente de água dentro de um cano, e tentar prosseguir a analogia encontrando paralelos entre a diferença de pressão ao longo do cano e diferença de voltagem ao longo do fio, o papel similar da viscosidade e da resistência elétrica etc. [10-11]. Precisamos enfatizar aqui que este tipo de analogia pedagógica não é o tema deste trabalho e não corresponde ao que queremos dizer por "metáfora científica".

Existe também um processo de criação de termos e mesmo conceitos científicos a partir do uso metafórico de vocábulos da linguagem comum. Ainda usando o exemplo anterior, os termos "carga", "corrente", "fio", "pressão", "resistência", "campo" etc. são etimologicamente anteriores ao seu uso científico, e tal uso se origina de uma ampliação do sentido original por via metafórica e analógica (e posterior definição de um novo sentido técnico literal). Também não chamaremos tais expressões de "metáforas científicas", mas sim de "metáforas comuns", ou seja, metáforas da linguagem comum aplicadas à linguagem científica.

Outros exemplos de termos em Física construídos a partir de metáforas comuns seriam: barreira entrópica, relevo de energia, poço de potencial, ruído branco, paisagem rugosa, rede cristalina, árvore de Cayley, bacia de atração, buraco negro, supercordas, dinâmica de avalanches etc. Note-se que, em geral, o termo coloquial é um substantivo simples com forte apelo imagético/sensorial (barreira, relevo, poço, ruído, rede, árvore, bacia, buraco, cordas, avalanches etc.) e que o mesmo é adjetivado, qualificado ou complementado a fim de formar uma expressão da terminologia científica.

Observamos que, se o uso de palavras comuns na criação de um termo científico pode trazer vantagens (em visualização, memorização, heurística analógica etc.), ela também pode acarretar desvantagens por remeter a sentidos que fogem ao desejado. Exemplos clássicos são as palavras “aceleração”, “força”, “peso”, “trabalho”, “energia”, “calor”, cujos sentidos coloquiais interferem com o aprendizado do sentido técnico no ensino de física [12, 13, 6, 2, 14]. A fim de evitar este perigo, é comum a criação terminológica de neologismos com um mínimo de sentido metafórico: quark, próton, entropia, entalpia, fractal, quasar etc. Isso não significa que tais termos não possam, no futuro, ser transferidos metaforicamente para a linguagem comum, como parece já estar ocorrendo com os termos entropia (como metáfora para desordem) e fractal (como metáfora para organização em vários níveis).

### 2.2 Linguagem comum e metáforas científicas

Chamaremos de metáfora científica (ou, se for necessário uma maior distinção, de “metáfora cientificamente inspirada” – MCI) o processo inverso ao discutido na seção anterior, ou seja, onde expressões com sentido técnico bem definido e estabelecido se transformam em expressões coloquiais através de um uso basicamente metafórico. Ou seja, conceitos científicos bem conhecidos são mapeados, pelo discurso comum, para outros domínios mais abstratos a fim de descrever sistemas complexos (usualmente psicológicos, biológicos, sociais e econômicos). Mostraremos a seguir que este processo é surpreendentemente comum para termos de física e geometria, embora muitas vezes passe despercebido. Discutiremos mais tarde sua potencial relevância para a questão do ensino e divulgação científicas.

Já observamos que o processo de criação de expressões comuns e termos técnicos não é unidirecional: um termo técnico pode ser construído a partir de termos totalmente coloquiais (por exemplo, “Big Bang”, “mundo pequeno”, “buraco negro”, “efeito borboleta”, “túnel de minhoca”), adquirir um sentido muito bem definido dentro do contexto científico e retornar ao vocabulário comum carregando, parcialmente que seja, o novo significado adquirido e gerando metáforas cientificamente inspiradas.

Assim, consideraremos que haverá uma metáfora científica se uma expressão com sentido técnico científico for usada metaforicamente em outros contextos, quer essa expressão seja um neologismo terminológico quer tenha ela mesma uma origem metafórica a partir de palavras comuns, conforme o processo descrito no item 2.1.

### 2.3 Metáforas visíveis e invisíveis, derivadas e básicas

Algumas expressões metafóricas foram incorporadas há tanto tempo pela linguagem comum e mesmo pelo léxico oficial que perdem grande parte de seu impacto de novidade semântica, de reorganização criativa do pensamento. São chamadas de “metáforas mortas”, confundindo-se, em certo limite, com catacreses (expressões incorporadas ao léxico e usadas na falta de vocábulos específicos, por exemplo “folha de papel”, “pés da mesa” etc). Preferimos descreve-las como “metáforas invisíveis” porque concordamos com Lakoff e Johnson [1] que as mesmas ainda podem afetar o pensamento e o comportamento por difundirem e reforçarem metáforas cognitivas (definidas a seguir) ou mesmo visões ideológicas. Em contraste, chamaremos de "metáforas visíveis" (na literatura, "metáforas vivas") àquelas em que o uso da terminologia científica ainda produz um choque semântico consciente, gerando novas percepções, analogias e mudanças conceituais. Estas serão tratadas na seção 4.

Notamos que o uso de termos da geometria Euclidiana são comuns em metáforas invisíveis:

1. *Ponto* de vista
2. *Linha* de raciocínio
3. Traçar um *paralelo*
4. Analisar por outro *ângulo*
5. *Volume* de conhecimentos
6. *Plano* pessoal
7. *Círculo* de amizades
8. *Triângulo* amoroso
9. *Esfera* de influência
10. *Pirâmide* social

Este uso mais antigo e incorporado na linguagem comum pode se dever ao fato de que a geometria Euclidiana faz parte da educação acadêmica desde o mundo greco-romano, mantendo-se também proeminente no *trivium* medieval. Ou seja, os termos da geometria Euclidiana (em contraste com a geometria não-Euclidiana e a geometria fractal, por exemplo) sempre fizeram parte do vocabulário das classes educadas e tiveram tempo suficiente para se difundir socialmente a partir delas.

Uma explicação alternativa, que não exclui a observação anterior de que a educação limita o repertório conceitual, se refere ao fato de que os termos de geometria Euclidiana são fortemente icônicos, e expressões relacionadas à visão e à ótica formam uma rede de metáforas derivada de uma "metáfora epistemológica básica" [1,2] relacionando conhecimento com visão: "ver aonde um argumento quer chegar", "enxergar um problema", "esclarecer uma dúvida", "refletir sobre um assunto", "iluminar uma questão", "o foco do estudo", "um assunto com vários aspectos", "idéia luminosa", "visão ideológica", "visão de mundo", "quadro mental", "sob essa lente", "sob esse prisma" etc. A palavra *teoria* vem do grego *theoria* com os significados de contemplar e especular (no sentido de espelhar). A própria idéia Cartesiana de que devemos pensar usando "idéias claras e distintas" em vez de metáforas é ela mesma uma expressão metafórica.

Ou seja, as expressões metafóricas 1-4 acima seriam derivadas da metáfora cognitiva básica "conhecer = ver" na medida em que tais figuras geométricas permitem descrever o processo de pensamento em termos visuais. Cabe notar que nossa própria terminologia de metáforas "visíveis" e "invisíveis" deriva desta metáfora básica. Chamaremos tais expressões metafóricas intimamente relacionadas a uma metáfora cognitiva básica de "metáforas derivadas".

Uma segunda metáfora epistemológica básica é a do canal de idéias, ou seja, os pensamentos seriam como objetos ou talvez como um fluido, e a comunicação se faz por "meios de transporte" (canais) que "carregam" a informação, e a mesma pode ser "armazenada" em recipientes como livros e discos. O uso do termo geométrico em

"volume de conhecimentos" (metáfora 5 da lista acima) talvez se relacione com isso, embora o fato de que livros, por ocuparem espaço físico, são organizados em "volumes", também reforce a metáfora de que idéias possuem volume físico e ocupam espaço, que a informação pode ser "comprimida" etc. A Teoria da Informação de Shannon transforma essa metáfora básica do "canal para o fluido do pensamento" no termo técnico "canal de informação". Curiosamente, o próprio termo "pensar", em sua origem latina, é de origem física e se refere ao ato de "pesar (as idéias) em uma balança".

Já o uso disseminado de metáforas geométricas (metáforas 6-10 acima) para representar relacionamentos sociais sugere a existência de uma terceira metáfora básica, talvez de origem antropológica, entre relação social e proximidade física (distância espacial em Geometria). Metáforas derivadas da mesma metáfora básica seriam: "parente próximo", "primo distante", "amigo chegado", "manter a distância patrão-empregado", "centro das atenções", "isolamento psicológico", "ator periférico", "inclusão social" etc. Notamos também que, ao mesmo tempo que geométrica, a expressão "triângulo amoroso" também sugere o grafo mais simples usado na moderna teoria de redes sociais, onde a noção de distância entre duas pessoas é medida agora pelo número de pessoas intermediárias ou "graus de separação" [15]. Acreditamos que a teoria de redes sociais e redes complexas irá gerar inúmeras metáforas científicas para o uso comum nos próximos anos.

## 3 Física clássica e metáforas sociais

Assim como as metáforas geométricas, a maior parte das metáforas originárias de termos da física clássica também perderam sua novidade semântica e possuem agora mais um caráter de metáforas invisíveis e convencionais: "forças políticas", "equilíbrio de poder", "fonte de atrito", "tensão social", "magnetismo pessoal" etc. já quase não remetem aos sentidos originais. O uso de metáforas físicas (e também geométricas e biológicas) quase passa despercebido quando lemos em um comentário econômico que "*Quando os saques ocorreram, eles simplesmente* ***implodiram*** *a* ***pirâmide*** *de papel, espalhando a crise por meio de* ***contágios*** *bancários. O* ***colapso*** *do comércio* ***transmitiu o choque****, tanto que a Alemanha e o Japão sofreram o* ***impacto*** *inicial ainda maior que os Estados Unidos e a Espanha"* [16], onde uma vaga noção da lei de conservação do momento linear parece estar presente.

O uso de metáforas físicas nas ciências sociais é bem documentado, tendo sua origem na "Física Social" do século XIX ou mesmo antes [17, 18, 20]. A terminologia de várias ciências sociais, em especial a sociologia, a ciência política e a economia, incorporaram de forma metafórica termos originários especialmente da Física Newtoniana. Em especial, o Marxismo clássico e a Psicanálise inspiraram-se em várias metáforas mecânicas e termodinâmicas [17-18].

Hoje, o uso dessa terminologia é amplamente disseminado, especialmente na linguagem jornalística, onde encontramos com frequência termos mecânicos como "correlação de forças", "movimentos sociais", "deslocamento do centro de gravidade do poder", "fase do ciclo econômico", "turbulência financeira", "equilíbrio fiscal",

“pressão social”, “tensão social”, “inércia governamental” etc., além de termos relacionados com o modelo mental “hidrodinâmico” de Freud e outros termos de origem termodinâmica muito presentes em Marx e Engels [18].

Novamente, propomos duas explicações complementares para a onipresença de termos de Física Clássica no discurso social: em primeiro lugar, tais metáforas agora invisíveis refletem a presença compartilhada de uma metáfora básica, a sociedade ou a economia como uma grande máquina ou, pelo menos, como um sistema (Newtoniano) de muitos corpos interagentes. Reconhecer isso não significa que tais metáforas básicas sejam especialmente pobres ou limitadas: a segunda metáfora, a de um sistema de muitas partículas fortemente interagentes (no sentido da Física Estatística ou Teoria de Sistemas Dinâmicos) capaz de apresentar fenômenos coletivos emergentes, transições de fase, dinâmica de avalanches, caos multidimensional etc. é uma abordagem científica promissora para o entendimento (ou pelo menos reconceitualização) de vários fenômenos sociais [19-23]. Notamos finalmente que existem também influentes metáforas básicas de origem biológica em ciências sociais: a sociedade como um organismo ou a economia como um ecossistema.

A presença da metáfora básica “a sociedade é um sistema de muitas partículas interagindo através de forças” não explica, porém, o fato de que as expressões metafóricas utilizadas no discurso jornalístico tenham um caráter antiquado, restringindo-se à Mecânica, Hidrodinâmica e Termodinâmica clássicas. Como vimos, a metáforas físico-matemáticas poderão, em um futuro próximo, envolver termos como bifurcações, transições de fase, avalanches, atratores, “efeito borboleta”, itinerância caótica [19-24] etc. para descrever fenômenos mentais e sociais. É preciso levar em conta o lento processo de difusão social.

A explicação mais simples é a de que jornalistas e colunistas tiveram como formação básica uma literatura clássica de ciências sociais ou humanidades, onde as metáforas da Física Clássica são usuais por terem sido usadas abundantemente pelo Marxismo e pela Psicanálise. De forma complementar, o leitor para o qual estes meios de comunicação se dirigem provavelmente teve acesso apenas a essa terminologia de física clássica no ensino médio (atrito, força resultante, pêndulo, centro de gravidade, inércia, pressão, temperatura etc.), de modo que a familiaridade com tais termos facilita a leitura das expressões metafóricas jornalísticas. Uma discussão sobre o descompasso histórico entre novas tecnologias e as metáforas tecnológicas antigas usadas para descrevê-la pode ser encontrada em [9].

Como exemplo de uso não trivial de metáforas Newtonianas no Marxismo, reproduzimos uma ilustração que Engels faz visando esclarecer a relação entre vontade individual e conseqüência histórica na visão marxista clássica:

> *A História se faz ela mesma de tal maneira que o resultado final é sempre oriundo de conflitos entre muitas vontades individuais, cada uma das quais, por sua vez, é moldada por um conjunto de condições particulares de existência. Existem* ***inumeráveis forças que se entrecruzam, uma série infinita de paralelogramos de forças que dão origem a uma resultante****: o fato histórico. Este, por sua vez, pode ser considerado como o* ***produto de uma força*** *que, tomada em seu conjunto, trabalha inconscientemente e involuntariamente. Pois*

> *o desejo de cada indivíduo é frustrado pelo de outro, e o que resulta disso é algo que ninguém queria.*
>
> *Assim é que* ***a História se realiza como se fosse um processo natural*** *e está sujeita, também, essencialmente às* ***mesmas leis de movimento.*** *Mas, do fato de que as diversas vontades individuais — cada uma das quais deseja aquilo a que a impelem a constituição física dos indivíduos e as circunstâncias externas (sejam pessoais ou da sociedade em geral que, em última instância, são econômicas) — não atinjam o que querem, mas* ***se fundam numa média coletiva, numa resultante comum****,* ***não se deve concluir que o seu valor seja igual a zero****. Pelo contrário, cada uma dessas vontades individuais* ***contribui para a resultante*** *e, nesta medida, está incluída nela. Eu pediria ao senhor que estudasse mais a fundo esta teoria nas suas fontes originais e não em fontes de segunda mão. Marx raramente escreveu alguma obra em que ela não tivesse seu papel, mas especialmente o "18 Brumário de Louis Bonaparte" é um excelente exemplo de sua aplicação* (carta de Friederich Engels a Konrad Schmidt, 5/8/1890).

Embora representar vontades por vetores (às quais se aplicaria o princípio da superposição linear e a soma dada por paralelogramos de força) seja um exagero metafórico, Engels descreve um programa de física estatística aplicado à descrição histórica que não difere em princípio de abordagens recentes em "econophysics" e "sociophysics" [19-23].

## 4 Metáforas científicas no discurso jornalístico: um estudo de casos

Existem inúmeras metáforas de física clássica e algumas de física moderna e contemporânea circulando no discurso comum e no discurso jornalístico. Examinaremos em detalhe o uso metafórico de um termo de física clássica ("pêndulo") e um termo de física contemporânea ("buraco negro").

### 4.1 O pêndulo e suas metáforas

Uma busca em portais jornalísticos brasileiros para o termo "pêndulo" no período de 2003 a 2009 retorna 109 textos na Folha Online, 68 no Estado de São Paulo Online e 57 no Portal G1. Nesta busca foram excluídos os textos repetidos. Na figura 1 mostramos a frequência entre usos próprios e usos metafóricos do termo "pêndulo" nessas diversas fontes. A proporção de uso de metáforas é: Folha = 57%, Estado = 62% e G1 = 29%. O fato de que Folha e Estado apresentam quase a mesma proporção é interessante e talvez decorra da natureza similar desses dois jornais em termos de público e corpo de jornalistas. Em contraste, o Portal G1 parece conter uma maior proporção de *press releases* científicos e menos artigos de colunistas (que em geral usam metáforas para tornar seu texto mais atraente).

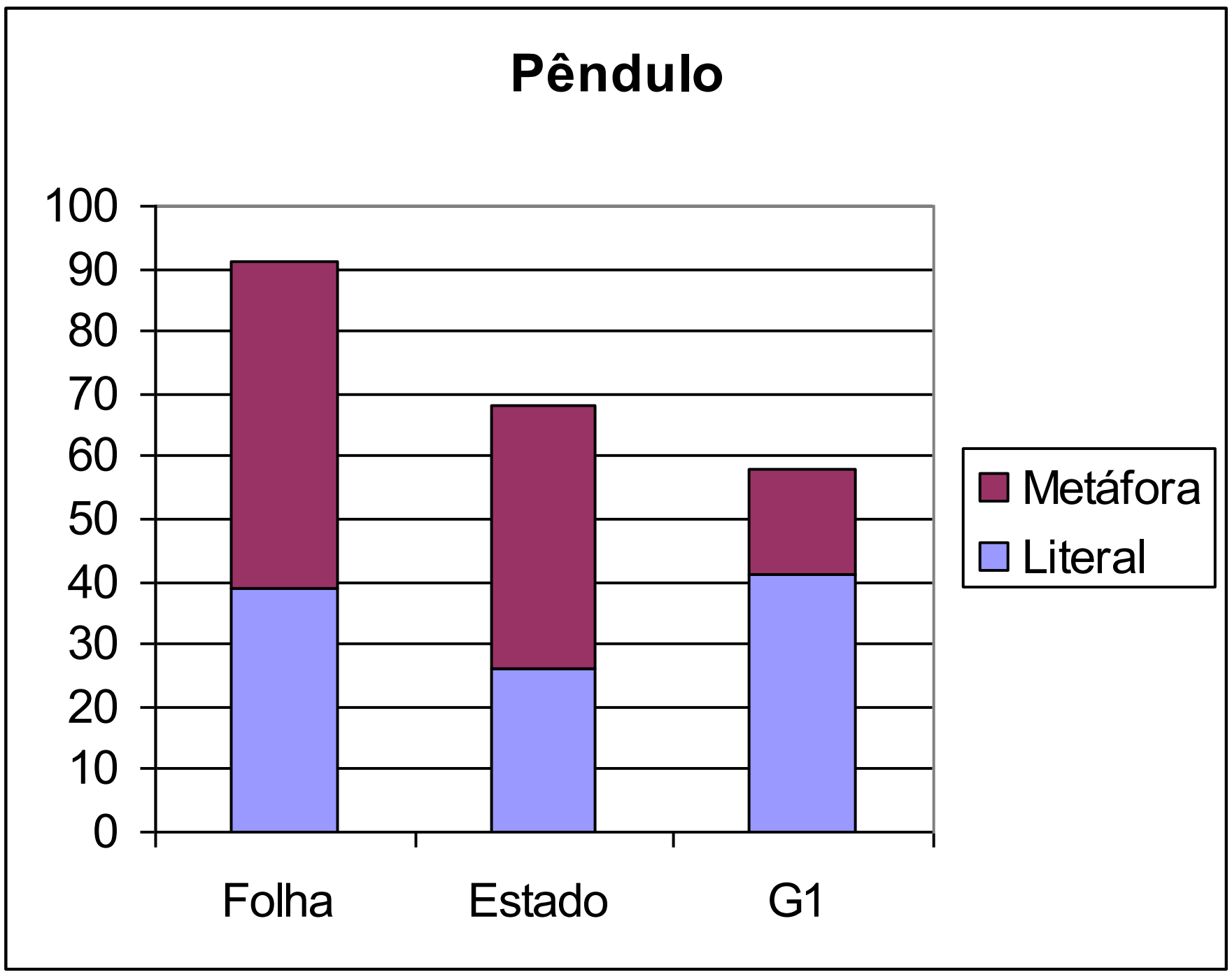

**Figura 1**: Número de usos literais e metafóricos da palavra "pêndulo" em nos portais jornalísticos Folha de São Paulo Online, Estado de São Paulo Online e Portal G1 no período 2002-2009.

Exemplificamos a seguir os usos metafóricos mais proeminentes. As metáforas não são excludentes mas possuem variações na ênfase que dão a diferentes aspectos do movimento pendular. Outros termos metafóricos de física e matemática também estão em negrito. Por economia de espaço, as citações não serão referenciadas, mas podem ser recuperadas usando-se máquinas de busca como o Google.

**a. Pêndulo como oscilação (quase) periódica:**

*Novo filme de Samira Makhmalbaf está no Festival do Rio*: Hana afirma que o **pêndulo** das relações profissionais para as familiares **oscila** com naturalidade entre eles.

*Um "Romeu e Julieta" novo, em vários sentidos*: Romeu e Julieta devem ser vistos como um **pêndulo** que **oscila** entre o amor e o ódio o tempo inteiro.

**b. Pêndulo como mudança recorrente, não necessariamente periódica:**

*Relação com Chávez afeta Kirchners: ...* construiu por muito tempo, particularmente até 2005, um vínculo com Hugo Chávez, sempre com muitas **oscilações**. Um **pêndulo** que **vibrava para o lado** do presidente venezuelano, mas que, às vezes, **pendia** com mais **intensidade** para o lado de Lula e...

*Federalismo truncado será o maior desafio*: Isso significa levar o **pêndulo** da descentralização até o fim e fazer a reforma tributária...examinar nossa história, vai notar um **pêndulo** entre **movimentos** de centralização e descentralização. A Constituição jogou o **pêndulo** para a descentralização, mas...

---

c. **Pêndulo como alternância:**

Grupo Sutil põe "véu da memória" em "Nostalgia": Trata-se de **estrutura** dramatúrgica que privilegia a **sobreposição de planos**, **pêndulo** do passado e do presente, invariavelmente sob o **ponto de vista** de um..

Paixões em **triângulo** de fogo e dor: Nesse **pêndulo** de frustrações e sonhos femininos, Ana Carolina de Lima **ilumina** a outra metade do jogo.

**d. Pêndulo como movimento entre dois pólos antagônicos:**

*Intelectual é fonte de pensamento independente no jornal*: ... ilusão de que essa seria uma saída para a **tensão** entre jornalistas e acadêmicos que tem caracterizado o **pêndulo** do jornalismo cultural nos últimos 30 anos.

*EUA mudarão política de clima, diz Gore:* Em política sempre há um **pêndulo balançando** da esquerda para a direita, e sempre que as pessoas percebem que o **pêndulo foi longe demais** para a direita elas o **puxam** de volta.

**e. Pêndulo como movimento cíclico:**

*Confiança do mercado se **dissolve** com **pressão** da dívida federativa*: Mesmo assim, a **ruptura** desse **círculo** -ou desse **pêndulo**- vicioso só deve vir com alguma grande notícia, como um nome para a nova...

*Retrato de uma artista:* ...se torna ruim, a cultura entra em **ebulição**. Quando o frenesi do dinheiro se acalma, a criatividade **emerge**. Isso é um **pêndulo**, um **ciclo natural** que vai permitir à moda **evoluir** e se **moldar** em novo formato.

**f. Pêndulo como indecisão entre alternativas:**

*O sucessor:* ...muito longo. Mas na hora da eleição, isso não tem um **peso** tão grande. Realmente há uma **oscilação**, uma espécie de **pêndulo**. Isso deve acontecer e os cardeais estarão pensando no bem da Igreja e não em como eleger uma pessoa que seria agradável...

*Deportações de imigrantes da Flórida cresceram em um ano*: ... dificuldade para conquistar eleitores hispânicos, principalmente de residentes da Flórida, considerado Estado-**pêndulo** (comumente indeciso entre republicanos e democratas).

---

**g. Pêndulo como movimento entre diversos polos:**

*D.O.M. é o melhor sul-americano, segundo 'Restaurant':* Parece que o **pêndulo** do gosto **oscila**, neste ano, entre os nórdicos e americanos, com **paradas** na Espanha e na Inglaterra.

**h. Pêndulo como sistema fora do ponto de equilíbrio:**

*Reconciliação interna é o grande desafio, diz político boliviano:* Para nós, a saída da Assembléia é conseguir um **equilíbrio**, que o **pêndulo** não vá ao Ocidente nem ao Oriente.

*Aula de yoga:* Descobri que a minha é como um **pêndulo**, às vezes está tudo **centradinho** e, de repente, pziiiin, lá se vai o **ímã desgovernado para a direita**. Concentro-me, corrijo as falhas, e o marcador retrocede obediente. Ele pára no meio, no **ponto perfeito**, mas não sei por que, pziiiin, novamente **escorrega para o outro lado**. Entrei na yoga para aprender a dominar o meu **pêndulo** interior.

É possível que esta metáfora caia em desuso, dado que hoje em dia poucas crianças terão contato com relógios de pêndulo (para uma visão alternativa, ver [9]). Nota-se que nessas aplicações, o pêndulo é uma metáfora inadequada por causa de seu movimento estritamente periódico. A inadequação da metáfora poderia ser superada caso o pêndulo caótico viesse a ser melhor conhecido como exemplo paradigmático de movimento recorrente porém não periódico [26]. A referência, na última metáfora, a um pêndulo que contém um imã poderia ser um reflexo de um contato da autora com tal aparato, hoje vendido em lojas de artefatos de decoração e curiosidades e apresentado em vídeos na internet.

### 4.2 Buracos Negros e suas metáforas

Uma busca para o termo “buraco negro” nos anos de 2008-2009 retorna 52 textos na Folha de São Paulo Online, 117 no Estado de São Paulo Online e 105 no Portal G1. Devemos notar que este período é bem menor que o período de 2002 a 2009 onde foram coletadas as expressões envolvendo o “pêndulo”, porém retorna um número similar de textos: 216 para “pêndulo” e 274 para “buraco negro”.

Na figura 2 mostramos a proporção entre usos próprios e usos metafóricos do termo “buraco negro”. Nota-se um uso menos frequente pela Folha (que aparentemente tem uma cobertura menor de notícias científicas que os outros portais). Entretanto, a proporção de uso metafórico é similar ao do portal do Estado de São Paulo e, novamente, uma diferença editorial em relação ao portal G1: Folha = 54% de metáforas, Estado = 46% e G1 = 35%. Essa alta proporção de uso metafórico mostra como as MCI podem ser extremamente comuns, às vezes superando a frequência do uso literal.

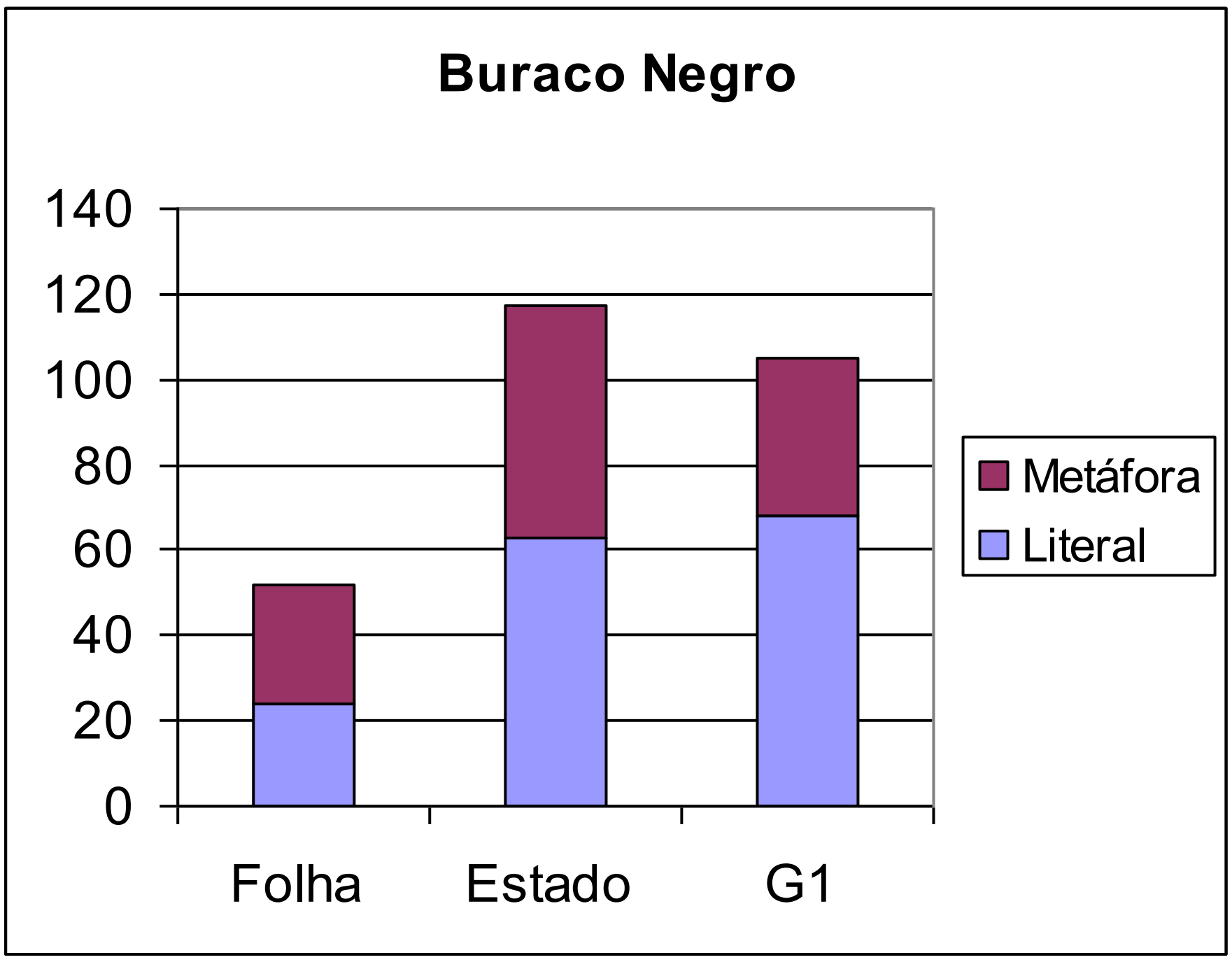

**Figura 2** Número de usos literais e metafóricos da expressão “buraco negro” nos portais jornalísticos Folha de São Paulo Online, Estado de São Paulo Online e Portal G1 no período 2008-2009.

Os sentidos metafóricos mais comuns e que contemplam diferentes aspectos do conceito científico de buraco negro são arrolados e exemplificados a seguir:

**a. Buraco negro como sumidouro que engole e faz desaparecer:**

*Europa pede provas de que presos de Guantánamo não são ameaça à segurança:* ... Irlanda e Suíça também disseram estar dispostos a receber os prisioneiros que, graças a um "*buraco negro*" legal criado pelos EUA, ficam anos presos sem acusação formal ou ...

*Twitter implementa opção de listas para facilitar organização:* ... No que diz respeito ao ambiente corporativo, o uso desses sites está se tornando claramente um **buraco negro** para a produtividade";, disse Philip Wicks ao jornal "Financial Times" ...

*Alegria do futebol*: ...nosso, em que ainda surgem talentos, apesar de tudo. Hoje esses jovens **brilham** uma, duas temporadas, e somem no **buraco negro** dos euros. Mas encantam enquanto passam, como **cometas**. Manda a sabedoria que aproveitemos a alegria enquanto dura...

*Frases*: "Não vou injetar dinheiro público em um buraco negro" Barack Obama, presidente dos EUA, prometendo não desperdiçar verba oficial.

**b. Buraco negro como depressão econômica profunda:**

*Sob efeito de crise nos EUA, Bovespa fecha com queda de 3,19%: ...* ;**derretimento**" (a **quebra**) do sistema financeiro americano --o que agiria como uma espécie de "**buraco negro**" na economia global, **arrastando** outros países para níveis mais agudos da ...

*Ventos pedem isonomias*: ...Desoneram-se tributos para investimentos produtivos no exterior, pois arrecadar menos é um esforço para deter o **buraco negro** da deflação.

*Todo dia é 11 de Setembro*: ...financeiro. Nessas horas, se alguém diz que o **universo** está prestes a ser **sugado** por **força gravitacional** para um **buraco negro** em Wall Street, quem tem dinheiro na bolsa se atira pra ver se salva o seu. E aí, meu amigo, é um 11 de Setembro...

**c. Buraco negro como depressão psicológica profunda:**

*Em biografia, goleiro Buffon revela ter sofrido de depressão*: ... nesta quinta-feira pelo jornal "La Stampa", Buffon, 30, conta como passou por um "**buraco negro**" durante seis meses. "Não estava satisfeito com minha vida e com o futebol e, ...

*Uma nação dividida pela raiva*: Daquele momento em diante, Britney se meteu num **buraco negro**. É um desastre após o outro.

**d. Buraco negro como lugar ou situação de difícil saída:**

*A educação dos sem-futuro*:...pão que manterá fingidamente de pé os que já caíram? Que desafio podem ter crianças e adolescentes para sair do **buraco negro** da ignorância, para estudar e aprender, ...

*Ipatinga perde para o guarani e entra na zona de rebaixamento*: ...vez nas trevas, Gustavo tentou de direita, mas acertou mesmo com a esquerda: 2 a 0 Guarani. O Ipatinga entrou no **buraco negro** da primeira divisão: é o penúltimo colocado.

**e. Buraco negro como pessoa ou objeto importante e que atrai:**

*Em festa de 50 anos, Zeca lota a Cidade do Samba*: ...Estava lá Lulu Santos, o roqueiro polêmico, para prestigiar o samba popular. 'Como não estar aqui. Zeca é o **buraco negro da galáxia**. Ele é tudo isso que **atrai** essa vida patrocinada pela alegria. Ele é tudo.

*Europeu ou americano, aqui é xis-queique*: O cheesecake do PJ Clarke's leva uma massa bem fininha e vai para a geladeira sem a calda. É um **buraco negro** de cremosidade, usando uma definição do crítico da revista americana Gourmet Francis Lam.

**f. Buraco negro como região do espaço não visível e da qual não se obtém informação:**

*Falta observação climática na Amazônia, alertam Nações Unidas*:...observação climática na Amazônia e que ainda existem "**buracos negros**" na cobertura de dados sobre a região. No caso do Brasil, Hinsman avalia que há "**buracos negros**" em termos de informação sobre a Amazônia...

*Voo Rio-Paris da Air France enfrenta emergência sobre Atlântico*: ... de Dacar, no Senegal, que é o responsável pela navegação aérea na região, conhecida como "**buraco negro**" pela falta de cobertura de radares aéreos.

**g. Buraco negro como lugar onde a informação é perdida ou esquecida:**

*Especialistas apoiam ideia mas temem repressão:...*Vieira de Oliveira, do Instituto Fernand Braudel. Hoje, os casos de violência registrados nas escolas entram no **buraco negro** dos boletins de ocorrência.

O luto no futebol: ...de dar um pitaco nessa matéria. Em casos parecidos, em vez de evitar a lembrança da derrota, de relegá-la ao **buraco negro** do esquecimento, acho que os jogadores deveriam discuti-la a fundo, ver e rever o teipe do jogo fatídico várias...

Novamente, observamos que o uso metafórico nem sempre respeita as propriedades do sentido literal original. Em particular, embora difícil, é possível sair de buracos negros metafóricos. Em todo caso, o termo parece ter sido rapidamente adotado por causa de sua força icônica, seu impacto conceitual: ao contrário dos termos metafóricos "depressão" e "poço sem fundo" aplicados a fenômenos econômicos e psicológicos, um "buraco negro" é mais ativo, ele atrai, suga, segura, impede a saída. Ou seja, velhas metáforas são substituídas por uma mais expressiva e descritiva, criada a partir da apropriação pelo leigo de um termo técnico da astrofísica.

---

## 5 Sugestões para a divulgação científica e o ensino formal de ciências

Termos científicos tanto se originam como, posteriormente, se transferem para a linguagem comum através de um uso metafórico. A partir da observação da onipresença de metáforas cientificamente inspiradas (visíveis ou invisíveis), gostaríamos de levantar questões e fazer algumas sugestões para a divulgação científica e o ensino de ciências.

### 5.1 Além de uma divulgação científica cientocêntrica: difusão de conceitos e vocabulário científicos como fator de ampliação de redes metafóricas cognitivas.

As mais variadas justificativas para a divulgação científica e a popularização da ciência têm sido apresentadas ao longo do tempo. Entre elas destacamos:

a) As descobertas científicas fazem parte da herança cultural comum da humanidade e constituem uma grande aventura intelectual. Assim como as artes, deveriam ser acessíveis a todas as pessoas;
b) A divulgação científica estimula o despertar de jovens talentos científicos;
c) Os cientistas precisam dar conta ao público dos resultados de suas pesquisas, dado que em grande parte elas são financiadas pelo Estado;
d) A cultura científica, por enfatizar o pensamento lógico e o ceticismo, estimulariam o pensamento crítico na população;
e) Grandes problemas da sociedade atual relacionados com fontes de energia, doenças infecciosas, engenharia genética, novas terapias, redes de informação etc. possuem forte componente científico e o debate democrático sobre tais problemas necessita de cidadãos razoavelmente informados cientificamente.

Acreditamos que todas essas razões têm o seu mérito, porém ao mesmo tempo são cientocêntricas, ou seja, atendem primariamente aos interesses da comunidade científica e secundariamente ao público (mesmo os últimos dois itens podem ser reformulados para que isso fique mais evidente).

A visão que defendemos neste trabalho pretende ser um pouco menos cientocêntrica: se a teoria cognitiva das metáforas estiver correta, então nossa capacidade de expressão e mesmo de pensamento é refém do repertório de metáforas a nossa disposição. Sendo que as grandes metáforas básicas cognitivas estão relacionadas com nosso corpo interagindo com o ambiente físico, as metáforas físicas e biológicas se tornam onipresentes em nosso discurso, mesmo sobre temas sociais, políticos ou filosóficos.

Parafraseando Engels que disse que "todo cientista é escravo de um filósofo morto" [18], podemos defender igualmente que "todo filósofo – e talvez toda pessoa comum – é escrava de um cientista morto". Usamos em nosso discurso diário uma profusão de metáforas mecânicas simples para descrever a História e sistemas socioeconômicos complexos (o exemplo da metáfora "pendular" exemplifica isso). Nosso repertório metafórico não apenas limita nossa capacidade de falar sobre tais sistemas, mas afeta nossa maneira de concebê-los e interagir com eles.

A moderna divulgação científica, via um *spin-off* não intencional, acaba gerando novas metáforas sociais e psicológicas que podem ser mais ricas, mais descritivas, mais apropriadas que as velhas metáforas do mecanicismo dos séculos XVIII e mesmo da Termodinâmica do século XIX [17]. Por exemplo, conceber o processo histórico como uma dinâmica fractal de avalanches de eventos, onde tanto a ação individual como os movimentos coletivos de todos os tamanhos são importantes [19,23], é um avanço frente a concepções simplistas como a História Whig de grandes personagens (metáfora do "indivíduo heróico") ou uma História determinada apenas por conflitos de classes (metáfora da "política como guerra"). Lembremos que a questão no discurso social amplo não é escolher entre um pensamento literal versus um metafórico, mas sim entre metáforas novas, mais complexas e descritivas, e metáforas antiquadas e muitas vezes inadequadas.

Acreditamos que esse papel da divulgação científica, até agora pouco estudado ou mesmo negligenciado, constitui um aspecto menos cientocêntrico da mesma: em vez de um conteúdo científico literal a ser absorvido por receptores sujeitos a ruído, o público apropria novos conceitos científicos e cria metáforas cientificamente inspiradas a serem usadas como ferramentas de pensamento e expressão: não apenas o meio é a mensagem, mas a metáfora é a mensagem.

### 5.2 Relação com os estudos de pré-concepções científicas

#### Metáforas como difusoras de vocabulário científico

Em relação à educação e divulgação científicas, o uso de metáforas científicas por jornalistas, colunistas, intelectuais etc. produzirá um resultado ambíguo. De um lado, haverá uma difusão de certo vocabulário que poderia facilitar a aprendizagem informal de certos conceitos [2]: por exemplo, dado que metáforas envolvendo a palavra "energia" são muito mais comuns que a palavra "entropia", é possível que isso contribua para que se possa pelo menos falar sobre energia no discurso comum, o que não é o caso do termo entropia.

Por outro lado, o uso metafórico implica que o sentido técnico não está sendo necessariamente respeitado. Ou seja, embora o uso metafórico contribua para que um termo científico seja mais usado ou conhecido pela população, também contribui para que tais termos sejam usados de forma equivocada no contexto técnico. Os sentidos metafóricos poderiam interferir com a aprendizagem ou a fixação do sentido técnico do termo no contexto da aprendizagem formal [13,25]. O caso da palavra energia seria então um exemplo clássico tanto do efeito facilitador como de interferência produzidas por metáforas cotidianas [2, 14].

#### Deteção e exame crítico de metáforas científicas invisíveis: um exercício de ensino formal para prevenção de pré-concepções científicas.

É plausível que muito da assim chamada estabilidade ou resiliência das "pré-concepções científicas equivocadas" estudadas na literatura [26, 12-14] se deve ao fato de que os termos científicos correspondentes são homônimos a termos da linguagem comum ou aplicados metaforicamente na linguagem comum. Afinal, estudantes não apresentam "pré-concepções equivocadas" sobre "momento angular", "momento de inércia", "dipolo", "entalpia", "geodésica" ou outros termos incomuns. Acreditamos que se os estudantes e professores reconhecerem previamente, através de exercícios, a onipresença das metáforas cientificamente inspiradas (MCI) no discurso comum, a discriminação e o aprendizado do sentido literal técnico e a prevenção de equívocos científicos serão imensamente facilitados. Estas afirmativas podem ser estudadas empiricamente.

Para a prevenção desses efeitos de interferência, propomos que a natureza e a ubiquidade das MCI, especialmente as derivadas de metáforas cognitivas básicas, seja reconhecida e enfrentada ativamente. Para isso sugerimos dois exercícios que podem ser realizados como trabalhos extra-classe.

**Exercício I – Examinando o uso metafórico de termos técnicos. Neste exercício o procedimento é similar ao que realizamos na seção 4.**

- **Detectando termos físico-matemáticos na linguagem cotidiana:** o estudante escolhe um dado termo técnico (por exemplo, "centro de gravidade", "calor" ou "energia") e examina seus usos metafóricos. Isso em pode ser feito usando-se máquinas de busca genéricas (Google, Bing, etc.), mas sugerimos o uso de ambientes mais controlados como portais jornalísticos.
- **Examinando o significado científico original dos termos:** isso pode ser feito a partir do livro-texto ou de dicionários/enciclopédias científicas confiáveis disponíveis na internet.
- **Detectando o uso metafórico e classificando metáforas:** o estudante deve separar expressões onde o termo científico é usado tecnicamente de outras expressões onde este é usado metaforicamente. Deve considerar se a metáfora linguística é ainda visível ou se já se tornou invisível. Se possível, a elaboração de um mapa conceitual que ligue as expressões metafóricas com as metáforas cognitivas básicas das quais foram derivadas, na forma de uma rede de metáforas com nodos e ligações, seria um exercício interessante.
- **Avaliando a qualidade do uso metafórico de termos de física (ou outras disciplinas):** finalmente, o estudante deverá opinar sobre a qualidade do mapeamento metafórico, ou seja, quais aspectos do sentido técnico do termo se mantém na metáfora, e quais foram claramente violados.

**Exercício II - Detectando metáforas básicas e metáforas invisíveis**

No exercício anterior, metáforas usando um mesmo termo são buscadas em diferentes textos. Neste exercício, escolhe-se apenas um texto rico em metáforas e propõe-se que as mesmas sejam localizadas e examinadas. Como exemplo, fazemos a análise do seguinte texto:

*É a ideologia que produz esse* ***efeito de evidência****, e de unidade, sustentando sobre o já dito os sentidos institucionalizados, admitidos como 'naturais'. Há uma parte do dizer, inacessível ao sujeito, e que fala em sua fala. Mais ainda: o sujeito toma como suas as palavras da voz anônima* ***produzida*** *pelo interdiscurso (memória discursiva). Pela ideologia, se naturaliza assim o que é* ***produzido*** *pela história: há transposição de certas* ***formas materiais*** *em outras, isto é, há simulação (e não* ***ocultação*** *de 'conteúdos') em que são construídas* ***transparências*** *(com se a linguagem não tivesse sua* ***materialidade****, sua opacidade) para serem interpretadas por* ***determinações*** *históricas que aparecem como evidências empíricas.*

*Redefinindo, assim, a ideologia discursivamente, podemos dizer que não há discurso sem sujeito nem sujeito sem ideologia. A ideologia, por sua vez, é interpretação de sentido em **certa direção**, **direção determinada** pela relação da linguagem com a história em seus **mecanismos** imaginários. A ideologia não é, pois, **ocultação** mas **função da relação necessária** entre a linguagem e o mundo. Linguagem e mundo se **refletem**, no sentido da **refração**, do **efeito** (imaginário) **necessário** de um sobre o outro. Na verdade, é o **efeito** da **separação** e da **relação necessária** mostrada nesse mesmo **lugar**. Há uma contradição entre mundo e linguagem e a ideologia é **trabalho** desta contradição. Daí a necessidade de distinguirmos entre a **forma** abstrata (com sua **transparência** e o efeito de literalidade) e a **forma material**, que é histórica (com sua **opacidade** e seus equívocos)"* [27].

Neste texto é fácil detectar as duas metáforas epistemológicas básicas de "conhecer = ver" (*transparências, ocultação, opacidade, refletem, refração, forma*) e "idéias = objetos materiais espacialmente localizados que podem ser deslocados e trabalhados" (*materialidade, trabalho, transposição, direção, forma material, conteúdo, separação, lugar, deslocamento*).

Existe também uma adesão às metáforas mecânicas deterministas típicas da Mecânica Clássica (*determinações históricas, produzido, determinada pela relação, mecanismos, função necessária, efeito, efeito necessário, relação necessária*). Uma visão probabilista da realidade, onde o acaso teria papel importante, está totalmente ausente. Assim, fazendo uma análise das metáforas usadas neste texto ficam evidentes a onipresença dos termos de Óptica Geométrica (às vezes impropriamente usados, como no caso do termo *refração*) e de Mecânica Clássica, explicitando-se metáforas cognitivas inconscientes talvez relacionadas a uma ideologia mecânica determinista.

## 6. Conclusões e Perspectivas

Todos usamos metáforas: não apenas como recurso expressivo, para tornar nosso discurso mais comunicativo ou icônico, mas sim porque o pensamento, o ato da cognição, são profundamente metafóricos e usamos metáforas para fazer o mapeamento entre domínios-origem concretos para domínios-alvo abstratos. De forma mais radical, talvez sejamos prisioneiros das metáforas que dispomos para nos expressar, as "metáforas pelas quais vivemos" [1] e até morremos.

Ao enriquecer o repertório conceitual de uma população, a educação e a divulgação científicas inevitavelmente irão produzir o surgimento de novas metáforas no discurso comum, especialmente àquelas relacionadas à tentativa de descrição de sistemas complexos como os sistemas sociais e econômicos. Esse processo não precisa ser visto como uma deturpação de conceitos científicos, mas sim como uma atualização e substituição de metáforas científicas antigas, ultrapassadas e invisíveis por outras mais descritivas, visíveis e semanticamente mais ricas. O objetivo deste trabalho foi chamar a atenção para este papel positivo, embora negligenciado, da educação e divulgação científicas.

As metáforas científicas veiculam tanto metáforas cognitivas básicas, por vezes ideológicas, como novas concepções capazes de mudar nossa visão de mundo. Estar atentos a elas em nossa fala e nossa escrita, tornar visíveis e criticar as metáforas invisíveis que assumimos tacitamente, e que podem constranger e limitar nosso pensamento e ação, promete ser um interessante exercício para educadores, divulgadores de ciência e estudantes.

**Agradecimentos:** O.K. agradece discussões com Roberto M. Takata, Antônio Carlos Roque da Silva, Luciano Bachmann e comentários de Mauro Rebelo, Lacy Barca e Tatiana Nahas. Os autores agradecem o apoio do CNPq.

**Bibliografia**